\documentstyle[twocolumn,aps,epsf,floats]{revtex}
\newcommand{\be}{\begin{equation}}
\newcommand{\ee}{\end{equation}}
\newcommand{\bea}{\begin{eqnarray}}
\newcommand{\eea}{\end{eqnarray}}

\begin{document} 

\twocolumn[\hsize\textwidth\columnwidth\hsize\csname %
@twocolumnfalse\endcsname

\title {SIN and SIS tunneling in cuprates.}
\author{Ar. Abanov and Andrey V. Chubukov}
\address{
Department of Physics, University of Wisconsin, Madison, WI 53706}
\date{\today}
\draft
\maketitle
\begin{abstract}
We calculate the SIN and SIS tunneling conductances
for the spin-fermion model. 
We argue that at strong spin-fermion coupling, relevant to cuprates, both conductances have dip features near the threshold frequencies when
 a tunneling electron begin emitting propagating spin excitation. 
We argue that  
the resonance spin frequency 
measured in neutron scattering can be inferred from the tunneling data by
analyzing the derivatives of SIN and SIS conductances.
\end{abstract}
]

\narrowtext
The electron tunneling experiments are powerful tools to study the spectroscopy of superconductors. These experiments 
%are usually performed using
%lock-in amplifier technique and 
measure the dynamical conductance
$dI/dV $ through a junction as a  
function of applied voltage $V$ and temperature~\cite{tunn,tunn2}.
For  superconductor-insulator-normal metal (SIN)
 junctions, the measured dynamical conductance is proportional to the 
 electron density of states (DOS) in a superconductor
$N(\omega) = -(1/\pi) \int d k ~Im G(k, \omega)$ at $\omega =eV~\cite{mahan}$.
For superconductor-insulator-superconductor (SIS) junctions,
the conductance 
$dI/dV \propto G(\omega = eV)$, where
%\begin{equation}
$G(\omega) = \int^{\omega}_0 d \Omega N(\omega -\Omega)~ \partial_\Omega
N(\Omega)$
%\label{sis}
%\end{equation}
is proportional to the derivative over voltage of the
convolution of the two DOS~\cite{mahan}.  

For conventional superconductors, 
the tunneling experiments 
have long been considered as one of the 
most relevant ones for the verification of the 
phononic mechanism of superconductivity~\cite{carbotte}. 
In this communication we discuss to
which extent the tunneling experiments on 
cuprates may provide the information about the 
pairing mechanism  in high-$T_c$ superconductors. 
More specifically, we discuss the implications of the 
spin-fluctuation mechanism of
high-temperature superconductivity on 
the forms of SIN and SIS dynamical conductances.

The spin-fluctuation mechanism  implies that
 the pairing between electrons is 
mediated by the exchange of their collective
spin excitations peaked at or near the antiferromagnetic momentum $Q$. 
This mechanism yields a $d-$wave superconductivity~\cite{dwave}, and
explains~\cite{expl,ac} a number of measured features in 
superconducting 
cuprates, including the peak/dip/hump features in the ARPES data 
near $(0,\pi)$~\cite{arpes}, 
and the resonance peak below $2\Delta $ in the inelastic neutron 
scattering data~\cite{neutrons}.
Moreover, in the spin-fluctuation scenario, the 
ARPES and neutron features are related:
the peak-dip distance in ARPES equals the resonance 
frequency in the dynamical spin susceptibility~\cite{ac}.
This relation has been experimentally verified in 
optimally doped and underdoped $YBCO$
and optimally doped $Bi2212$ materials~\cite{norman}.
Here we argue that  
the resonance spin frequency  can also 
be inferred from the tunneling data by
analyzing the derivatives of SIN and SIS conductances.

The SIN and SIS tunneling experiments have 
been performed on $YBCO$ and 
$Bi2212$ materials~\cite{tunn,tunn2}. At low/moderate frequencies, both
 SIN and SIS conductances display a
behavior which is generally expected in a
$d-$wave superconductor: SIN conductance is linear in
voltage for small voltages, and has a peak at $eV = \Delta$ where $\Delta$ is
the maximum value of the $d-$wave gap~\cite{tunn}, 
while SIS conductance is quadratic in
voltage for small voltages, and has a near discontinuity at $eV = 2 \Delta$~\cite{tunn2}.
These features have been explained by a weak-coupling theory, 
without specifying
the nature of the pairing interaction~\cite{maki}. 
However, above the peaks, both
SIN and SIS conductances have extra dip/hump features which become visible at
around optimal doping, and grow
with underdoping~\cite{tunn,tunn2}. We argue that these features are 
sensitive to the type of the pairing interaction
 and can be explained in the spin-fluctuation theory.
%We show that in the spin fluctuation scenario, 
%the location of the dip is related to the resonance frequency in the dynamical
%spin susceptibility.   

As a warm-up for the strong coupling analysis, consider first 
SIN and SIS tunneling in a $d-$wave superconductor in the weak coupling limit.
In this limit, the fermionic self-energy is neglected, and the 
superconducting gap does not depend on frequency. For simplicity, we
consider  a circular Fermi surface for which $\Delta_k = \Delta \cos 2\phi$.

\begin{figure} [t]
\begin{center}
\leavevmode
\epsfxsize=3.3in 
\epsfysize=1.8in 
\epsffile{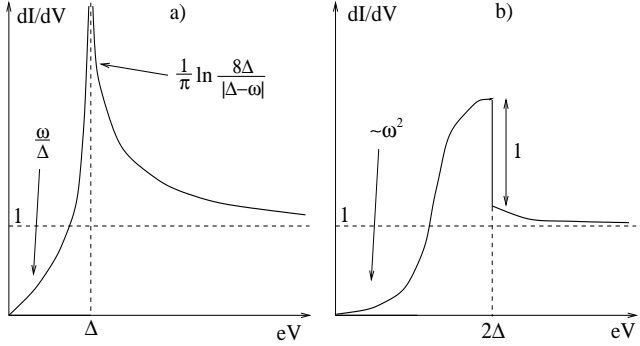}
\end{center}
\caption{The behavior of SIN and SIS tunneling conductances, $dI/dV$, in 
a $d-$wave BCS superconductor (figures a and b, respectively)}. 
\label{fig1}      
\end{figure}

We begin with the SIN tunneling. Integrating 
$G(k, \omega) = 
(\omega + \epsilon_k)/(\omega^2 - \epsilon_k^2 - \Delta^2_k)$ over 
$\epsilon_k = v_F (k-k_F)$ we obtain
\begin{eqnarray}
N (\omega  ) &=& Re~ \frac{\omega }{2\pi }\int _{0}^{2\pi }
\frac{d\phi }{\sqrt{\omega ^{2}-\Delta ^{2}\cos ^{2}(2\phi )}} \nonumber\\
&=&\frac{2}{\pi }\left\{ 
\begin{array}{ll}
K(\Delta /\omega ) & \mbox{for $\omega >\Delta $}\\
(\omega /\Delta )K(\omega /\Delta ) & \mbox{for $\omega <\Delta $}
\end{array}\right. ,                        \label{sin-gas}
\end{eqnarray}
where $K(x)$ is the elliptic integral.
We see that $N(\omega)
\sim \omega $ for $\omega \ll \Delta $ and diverges logarithmically as 
$(1/\pi)\ln (8\Delta/|\Delta -\omega|)$ for $\omega \approx \Delta$. At larger
frequencies, $N(\omega)$ gradually decreases to a frequency independent, normal
state value of the DOS, which we normalized to $1$. 
The plot of  $N(\omega)$ is presented in
Fig~\ref{fig1}a.

We now turn to the SIS tunneling.
Substituting the results for the DOS into $G(\omega)$ 
and integrating over $\Omega$, 
we obtain the result presented in 
Fig~\ref{fig1}b. At small $\omega$, $G(\omega)$
 is quadratic in frequency, which is an
obvious consequence of the fact that 
the DOS is linear in $\omega$. At 
 $\omega = 2\Delta$, $G(\omega)$ undergoes a finite jump. 
This discontinuity is related to the fact that near 
$2\Delta$, the integral over the two DOS  includes the region
$\Omega \approx \Delta$ 
where both $N(\Omega)$ and $N(\omega -\Omega)$ are  logarithmically singular, 
and $\partial_\Omega N(\Omega)$ diverges as $1/(\Omega-\Delta)$.
 The singular 
contribution to $G(\omega)$ from this 
region can be evaluated analytically and yields
\begin{equation}
%G^{sing}
G(\omega) = -\frac{1}{\pi ^{2}}P\int _{-\infty }^{\infty } 
%d x~
\frac{d x~\ln |x|}{x+\omega -2\Delta } 
%+\mbox{regular terms} \nonumber \\&=&
= -\frac{1}{2}\mbox{sign}(\omega -2\Delta )
%+\mbox{regular terms}
                      \label{sis-gas}
%\end{eqnarray}
\end{equation}
%where the regular terms are the contributions from energies  away
%from $\omega =\Delta$. 
We see that the amount of 
jump in the SIS conductance is a universal number which 
does not depend on $\Delta$.

The results for the SIN and SIS conductances in a $d-$wave 
gas agree with earlier studies~\cite{maki}. 
In previous studies, however, SIS conductance
was computed numerically, and the universality of the amount of the jump at
$2\Delta$ was not discussed, although it is clearly 
seen in the numerical data. 

We now  turn to the 
main subject of the paper  and discuss the forms of SIN and SIS
conductances for strong spin-fermion interaction.

We first show that the features observed in a gas are in fact quite general
and are 
present in an arbitrary Fermi liquid as long as the 
impurity scattering is weak. Indeed, in an arbitrary $d-$wave superconductor,  
\begin{equation}
N(\omega) \propto Im~\int  d \phi~ 
\frac{\Sigma (\phi ,\omega )}{(F^2 (\phi,\omega) - 
\Sigma^{2} (\phi ,\omega ))^{1/2}},
\label{eq}
\end{equation}
 where  $\phi $ is the angle along the Fermi 
surface, and $F (\phi, \omega)$ and 
$\Sigma (\phi, \omega)$ are the retarded 
anomalous pairing vertex and retarded fermionic
self-energy at the Fermi surface (the latter includes a bare $\omega$ term
in the fermionic propagator). 
 The measured superconducting gap  $\Delta (\phi)$ is a solution of 
$F (\phi,\Delta (\phi)) = \Sigma (\phi, \Delta (\phi))$.

In the absence of impurity scattering,
$Im \Sigma$ and $Im F$ in a superconductor both vanish at $T=0$ 
up to a frequency
which for arbitrary strong interaction  exceeds $\Delta$. 
The Kramers-Kronig relation then yields at low frequencies 
$Re \Sigma (\phi, \omega) \propto \omega$, $Re F(\phi, 
\omega) \propto (\phi - \phi_{node})$ where $\phi_{node}$ is a position of the
node of the $d-$wave gap. Substituting these forms into  
(\ref{eq}) and integrating over $\phi$ 
we  obtain $N(\omega) \propto \omega$ although the
prefactor is different from that in a gas. The linear behavior of the
DOS in turn gives rise to the quadratic behavior of the SIS conductance.

Similarly, expanding $\Sigma^2 - F^2$ near each of the maxima  of
the gap  we obtain 
$\Sigma^2 (\phi, \omega) -F^2 (\phi, \omega) \propto (\omega - \Delta)
+ B (\phi-\phi_{max})^2$, where $B>0$. Then
%\begin{eqnarray}
\begin{equation}
N (\omega ) \propto
Re\int \frac{d{\tilde\phi} }{\sqrt{B {\tilde\phi}^{2} + 
(\Omega -\Delta)}} \approx -\frac{\ln |\Omega -\Delta |}{\sqrt{B}}
%\mbox{regular terms}
\label{sin-strong}
%\end{eqnarray}
\end{equation}
This result implies that the SIN conductance in an arbitrary Fermi liquid 
still has a logarithmic
singularity at $eV = \Delta$, although its residue depends on 
the strength of the 
interaction. The logarithmical divergence of the DOS causes the discontinuity
in the SIS conductance by the same reasons as in a Fermi gas.

In the presence of impurities,  the logarithmical singularity
is smeared out, and the DOS acquires a nonzero value
at zero frequency  (at least, in the self-consistent $T-$matrix approximation~\cite{t-mat}). However, for small concentration of impurities, this 
affects the conductances only in narrow frequency regions near singularities
while away from these regions the behavior is the same as in the absence of impurities. 

\begin{figure} [t]
\begin{center}
\leavevmode
\epsfxsize=3.3in 
\epsfysize=1.8in 
\epsffile{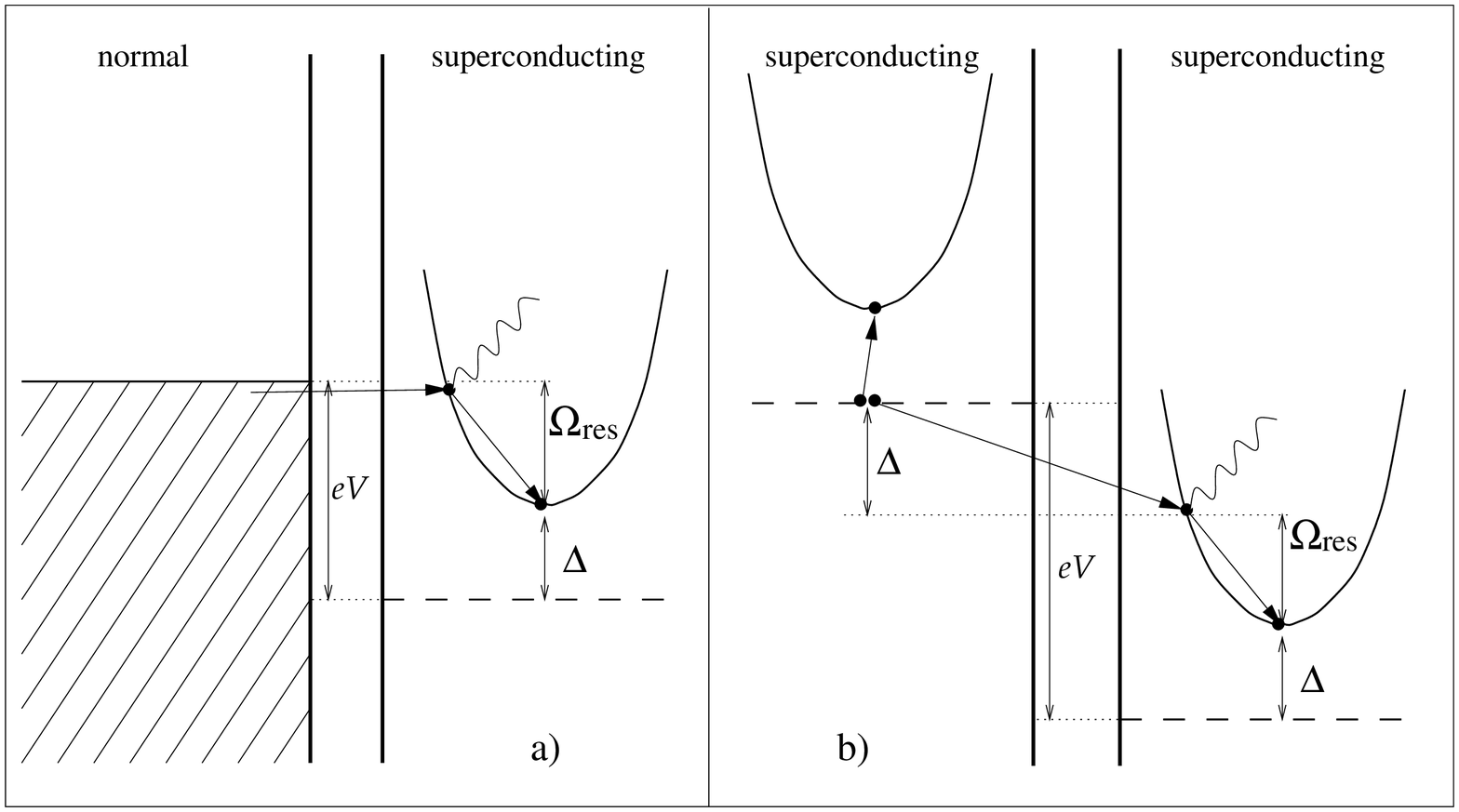}
\end{center}
\caption{The schematic diagram for the dip features in SIN and SIS tunneling
conductances (figures a and b, respectively). 
For SIN tunneling, the 
electron which tunnels from a normal metal can emit a propagating spin wave 
if the voltage $eV = \Delta + \Omega_{res}$ where $\Omega_{res}$ is the minimum
frequency for spin excitations. After emitting a spin-wave, the electron falls to the bottom of the band which leads to a sharp reduction of the current and produces a drop in $dI/dV$. For SIS tunneling, the physics is similar, but one first has to break an electron pair, which costs energy $2\Delta$.}
\label{fig2}      
\end{figure}

We now show that a strong spin-fermion
interaction gives rise extra features in the SIS and SIN conductances, not present in a gas.
The qualitative explanation of these features is the following.
At strong spin-fermion coupling, a $d$-wave superconductor
possesses {\it propagating}, spin-wave type  collective spin excitations 
near antiferromagnetic momentum $Q$ and at frequencies below $2\Delta$. 
These excitations
give rise to a sharp peak in
the dynamical spin susceptibility at a frequency $\Omega _{res} <2\Delta$
~\cite{neutrons}, 
and also contribute to the damping of 
fermions near hot spots (points at the Fermi surface separated by $Q$), where
the spin-mediated $d-$wave superconducting gap is at maximum. 
If the voltage for SIN tunneling is such that 
$eV=\Omega _{res}+\Delta$, then an electron which tunnels from the normal
metal, can emit a spin excitation and fall to the bottom of the band 
(see Fig.~\ref{fig2}a) 
loosing its group velocity. This obviously leads to a sharp 
reduction of the current and produce a drop in $dI/dV$. 

Similar effect holds for SIS tunneling. Here however 
one has to first break an 
electron pair, which costs the energy $2\Delta$. After a pair is broken, one of the 
electrons becomes a quasiparticle in a superconductor and takes an energy $\Delta$, while the other tunnels.
If $eV = 2\Delta + \Omega_{res}$, the electron which tunnels through a barrier
has energy $\Delta + \Omega_{res}$, and
can emit a
 spin excitation and fall to the bottom of the band. This again produces a
sharp drop in $dI/dV$ (see Fig.~\ref{fig2}b).

In the rest of the paper we consider this effect in more detail and make
quantitative predictions for the experiments. Our goal is to compute
$dI/dV$ for SIN and SIS tunneling for strong spin-fermion interaction.

The point of departure for our analysis is the set of  two
Eliashberg-type equations for  the fermionic
self-energy $\Sigma_\omega$, and the spin polarization 
operator $\Pi_\Omega$. The later is
 related to the dynamical spin susceptibility at the antiferromagnetic momentum by $\chi^{-1} (Q, \Omega) \propto 1- \Pi_\Omega$.
 The same set
 was used in our earlier analysis of the relation
between ARPES and neutron data~\cite{ac}.
 In Matsubara frequencies these equations  read
(${\tilde \Sigma}_{\omega_m} = i \Sigma ({\omega_m})$)
\begin{eqnarray}
{\tilde\Sigma}_{\omega_m} &=& {\omega_m} + \frac{3R}{8 \pi^{2}}~\int
\frac{{\tilde\Sigma}_{{\omega_m} +{{\Omega_m}}}}
{q_{x}^{2}+{\tilde\Sigma}^{2}_{{\omega_m} +{{\Omega_m}}}+F^2}
\frac{d {{\Omega_m}}}{\sqrt{q_{x}^{2}+1 -\Pi_\Omega }} 
\nonumber \\
%{\tilde\Sigma}_{\omega_m} &=& {\omega_m} + \frac{3R}{8 \pi}~\int
%\frac{{\tilde\Sigma}_{{\omega_m} +{{\Omega_m}}}}
%{\sqrt{{\tilde\Sigma}^{2}_{{\omega_m} +{{\Omega_m}}}+F^2}}
%\frac{d {{\Omega_m}}}{\sqrt{1 -\Pi_\Omega }} 
%\nonumber \\
\Pi_\Omega&=& \frac{1}{2}
\int \frac{d {\omega_m}}{\omega_{sf}}~\left(
\frac{{\tilde\Sigma}_{{{\Omega_m}} +{\omega_m}}~{\tilde\Sigma}_{{\omega_m}} + F^2}
{\sqrt{{\tilde\Sigma}^{2}_{{{\Omega_m}} +{\omega_m}}+F^2}~
\sqrt{\tilde\Sigma^{2}_{\omega_m}+F^2}}-1 \right).
\label{set2b}
\end{eqnarray}
This set is a simplification of the full set of Eliashberg equations
that includes also the 
equation for the anomalous vertex $F(\omega)$~\cite{acf}. 
As in~\cite{ac} we assume  that near  optimal doping,
the  frequency dependence of $F(\omega)$ is weak at $\omega
\sim \Delta$ relevant to our analysis,
and replace $F(\omega)$ by a
frequency independent input parameter $F$.
Other input parameters in (\ref{set2b}) 
 are the dimensionless  coupling constant
$R = {\bar g}/(v_F \xi^{-1})$ and
a typical spin fluctuation
 frequency 
$\omega_{sf} = (\pi/4) (v_F \xi^{-1})^2/{\bar g}$. They are
expressed in terms of
the effective spin-fermion coupling constant ${\bar g}$, the 
Fermi velocity at a hot spot $v_F$, and the
magnetic correlation length $\xi$.
 By all accounts, at
 and below optimal doping, $R \geq 1$~\cite{chubukov}, i.e., the system behavior falls into the strong coupling regime.
 
Strictly speaking,
 the set (\ref{set2b}) is valid near hot spots where $\phi \approx
\phi_{max}$. Away from hot spots $F (\phi)$ is reduced compared to 
$F$. We, however, will demonstrate that the
new features due to spin-fermion interaction are produced solely 
by fermions from hot regions. 
%Furthermore, there are indications that the
%tunneling matrix elements for SIS junctions are weighted towards hot regions~\cite{john}.
 
As in ~\cite{ac}, we consider the solution of (\ref{set2b}) for the
experimentally relevant case $F \gg R \omega_{sf}$ when the measured
superconducting gap $\Delta \sim F^2/(R^2 \omega_{sf}) \gg \omega_{sf}$.
In this situation, 
 at frequencies $\sim \Delta$, 
 fermionic excitations in the normal state are overdamped due to strong spin-fermion interaction. In a 
superconducting state, the form of the spin propagator is modified 
 at low frequencies because of the gap opening, 
and this gives 
rise to a strong feedback from superconductivity on the electron DOS.

More specifically, 
we argued in ~\cite{ac} that in a superconductor, 
$\Pi_\Omega$ at low frequencies $\Omega \ll 2\Delta$ 
behaves as $\Omega^2/\Delta$, i.e., collective spin excitations are undamped, 
propagating spin waves. This behavior is peculiar to a superconductor -- 
in the normal state, the spin excitations are completely overdamped.
The propagating excitations give rise to the resonance in $\chi (Q,\Omega)$
at $\Omega _{res}\sim (\Delta \omega _{sf})^{1/2} \ll \omega_{sf}$
where $Re\Pi (\Omega _{res})=1$~\cite{ac}.
This resonance accounts for the peak in neutron scattering~\cite{neutrons}.
% The condition $\omega_{sf} \ll \Delta$
%implies that $\Omega_{res} \ll \Delta$ and justifies the use of the 
%$\Omega^2$ form for $\Pi_\Omega$.

The presence of a new magnetic 
propagating mode changes the electronic self-energy for 
electrons near hot spots. In the absence of a propagating mode,
an electron can decay only if its energy exceeds $3\Delta$.
Due to resonance, an electron at a hot spot can emit a spin wave  already when its energy
exceeds $\Delta + \Omega_{res}$. It is essential that contrary to
a conventional electron-electron scattering,  this
 process gives rise to
a discontinuity in $Im \Sigma (\omega)$ at the threshold. 
Indeed, using the spectral representation to 
transform from Matsubara to real frequencies
in the first equation in (\ref{set2b}), integrating over momentum 
% fact that $\Sigma (\omega) \propto
%\omega$ at $\omega \sim \Delta$, 
and neglecting for simplicity unessential
$q^2_x$ in the spin susceptibility,
 we obtain for $\omega \geq \omega_{th} = \Delta + \Omega_{res}$
% and 
%using the spectral representation for the transformation from Matsubara to real frequencies
%\begin{equation}
%Im \Sigma (\omega) \propto 
%\int^\omega_0 d\Omega Im \frac{\omega- \Omega}{\sqrt{\Delta^2 - 
%(\Omega-\omega)^2}}~
%Im \frac{\Omega_{res}}{\sqrt{\Omega^2_{res}- \Omega^2}}
%\label{new1}
%\end{equation}
%The product in the integrand in  (\ref{new1}) is nonzero when 
%$\omega - \Omega > \Delta$, and $\Omega > \Omega_{res}$, i.e., when
%$\omega > \omega_{th} = \Delta + \Omega_{res}$. Expanding 
%the integrand at $\omega \geq \omega_{th}$  we
%obtain after simple manipulations
\begin{eqnarray}
Im \Sigma (\omega) &\propto& \int^{\omega-\Delta}_{\Omega_{res}} 
d \Omega \frac{1}{\sqrt{\omega - \Omega -
\Delta}}~\frac{1}{\sqrt{\Omega - \Omega_{res}}} \nonumber \\
&&\propto \int^{(\omega - \omega_{th})^{1/2}}_0 dx 
\frac{1}{\sqrt{\omega - \omega_{th} -x^2}} = \frac{\pi}{2},
\end{eqnarray}
We see that $Im \Sigma (\omega)$ jumps 
to a finite value at the threshold. This discontinuity is peculiar to
two dimensions.
By Kramers-Kronig relation, the 
discontinuity in $Im \Sigma$ gives rise to a logarithmical divergence of $Re
\Sigma$ at $\omega = \omega_{th}$.  This in turn 
gives rise to a vanishing spectral
function near  hot spots, and accounts for a sharp dip in the ARPES data~\cite{arpes}.

We now show that the singularity in
 $Re \Sigma (\omega)$ causes the singularity in the derivatives over voltages
of both SIN and SIS conductances $d^2I/dV^2$.
Indeed, near a hot spot, $F(\phi) = F (1- \lambda {\tilde \phi}^2)$ where
${\tilde \phi} = \phi - \phi_{max}$, and $\lambda >0$.
Then, quite generally, $Re \Sigma (\phi,
\omega) \propto \ln |\omega - \omega_{th} (\phi)|$  where 
$\omega_{th} (\phi) = \omega_{th} + C {\tilde \phi}^2$, and $C>0$.
 Substituting  
this expression into the DOS and differentiating over frequency, we obtain
after a simple algebra
\begin{eqnarray}
\frac{\partial N (\omega )}{\partial \omega } 
&\sim& -\int \frac{F^{2}(\phi )}{\Sigma^{3}(\phi ,\omega ) }
\partial_{\omega } \Sigma (\phi ,\omega )d\phi 
%+\mbox{regular terms} 
\nonumber \\
%&& \sim  \frac{1}{\ln ^{3}|\omega -\omega_{th}|}
%P\int \frac{d\phi }{C{\tilde\phi}^{2}+\omega_{th}-\omega }
%\nonumber \\
&&\sim  \frac{1}{\ln ^{3}|\omega - \omega_{th}|}
\frac{\Theta (\omega_{th}-\omega)}
{\sqrt{\omega_{th}-\omega}}, \label{sin-strong-spin}
\end{eqnarray}
where $\Theta(x)$ is a step function.
We see that $\partial N(\omega)/\partial\omega$ has a one-sided, square-root singularity
at $\omega = \omega_{th}$. 
Physically, this implies that the 
conductance drops when propagating
electrons start emitting spin excitations. 
Note that the typical $\phi$ which contribute to the singularity are small 
(of order $|\omega_{th} - \omega|^{1/2}$), which 
justifies our assertion that the singularity is confined to hot spots.

\begin{figure} [t]
\begin{center}
\leavevmode
\epsfxsize=3.3in 
\epsfysize=1.8in 
\epsffile{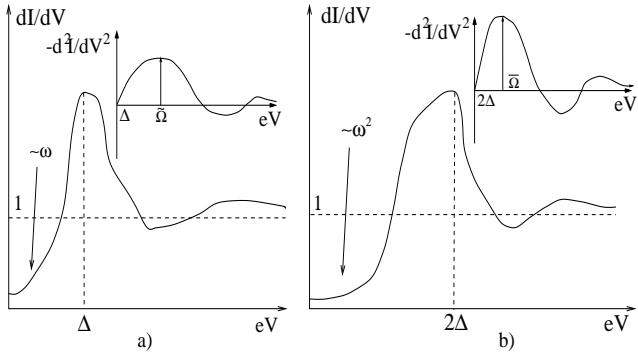}
\end{center}
\caption{The schematic forms of SIN (a) and SIS (b)
 tunneling conductances for strong spin-fermion
 interaction. We added small impurity scattering to soften singular features
related to the sharpness of the Fermi surface (see the text). The dip/hump features above the peaks are the strong coupling effects not present in a gas. 
The insets show the derivatives of conductances 
above $\Delta$ for SIN tunneling and $2\Delta$ for SIS tunneling.
 We argue in the text that 
these derivatives have maxima at voltages 
 $eV = {\tilde \Omega} = 
\Delta + \Omega_{res}$ for SIN tunneling and 
$eV = {\bar \Omega} = 2\Delta + \Omega_{res}$ for SIS tunneling, 
 where $\Omega_{res}$ is the resonance spin frequency measured in neutron experiments.}
\label{fig3}      
\end{figure}

The singularity in $\partial N(\omega)/\partial \omega$ is likely to give rise to a  dip 
in $N(\omega)$ at $\omega \geq \omega_{th}$. The argument here 
is based on the fact that 
if the angular dependence of $\omega_{th} (\phi)$ is weak
(i.e., $C$ is small), then $\Sigma (\omega_{th}) \gg F(\omega_{th})$, and 
$N(\omega_{th})$ reaches its normal state value with infinite negative
derivative. Obviously then,  at $\omega > \omega_{th}$, $N(\omega)$
 goes below its value
in the normal state and should therefore  have a minimum
at some $\omega \geq \omega_{th}$.
 Furthermore, at
larger frequencies, we solved (\ref{set2b}) 
perturbatively in $F(\omega)$ and found that $N(\omega)$ 
approaches a normal state value {\it from above}. This implies that besides a
dip,  $N(\omega)$ should also display a 
hump somewhere above $\omega_{th}$. The behavior of the SIN conductance is
schematically shown in Fig.~\ref{fig3}a.

Similar results hold for SIS tunneling.
The derivative of the SIS current, $d^{2}I/dV^{2} \sim 
\partial G(\omega)/\partial \omega$, is given by 
\begin{equation}
\frac{\partial G(\omega)}{\partial \omega} = 
 \int _{0}^{\omega}
\partial _{\omega}N (\omega -\Omega )
\partial _{\Omega }N (\Omega )d\Omega 
\label{sis2}
\end{equation}
Evaluating the integral in the same way as for SIN tunneling, we find 
a square-root singularity 
at $\omega = \omega^*_{th} = 2\Delta +\Omega _{res}$.
\begin{eqnarray}
\frac{d^{2}I}{dV^{2}} &\sim& -P\int_{0}^{\omega }
\frac{d\Omega }{\omega -\Omega-\Delta}~
\frac{1}{\ln ^{3}|\omega_{th}-\Omega |}
\frac{\Theta (\omega_{th}-\omega)}
{\sqrt{\omega_{th}-\omega}} \nonumber \\
&\sim &-\frac{1}{\ln^{3}|\omega^*_{th}-\omega |}
\frac{\Theta (\omega^*_{th} - \omega)}
{\sqrt{\omega^*_{th} - \omega}}
\label{sis-strong-spin}
\end{eqnarray}
 The singularity 
comes from the region 
where $\Omega \approx \omega_{th}$ and $\omega-\Omega \approx \Delta$, and 
both $\partial_{\omega} N (\omega -\Omega )$ and 
$\partial _{\omega }N (\omega )$ 
are singular.

Again, it is very plausible that the
singularity of the derivative causes a dip at a frequency $\omega \geq
\omega^*_{th}$, and a hump at even larger frequency. We stress, however, that 
at exactly $\omega^*_{th}$, the SIS conductance has an infinite
derivative, while the dip occurs at a frequency which is somewhat 
larger than $\omega^*_{th}$. The behavior of the SIS conductance is presented
in Fig~\ref{fig3}.

Qualitatively, the forms of conductances presented in Fig~\ref{fig3} agree with
the SIN and SIS data for YBCO and Bi2212 materials~\cite{tunn,tunn2}. Moreover,
recent SIS tunneling data for 
$Bi2212$~\cite{tunn2} indicate that the relative distance between the peak  and the dip ($\Omega_{res}/(2\Delta)$ in our theory)
 decreases with underdoping. More data analysis is however necessary to quantitatively compare tunneling and neutron data.

To summarize, in this paper we considered the forms of SIN and SIS conductances
both for noninteracting fermions, and for fermions which strongly interact with
their own collective spin degrees of freedom.
We argue that for strong spin-fermion interaction, 
the resonance spin frequency $\Omega_{res}$
measured in neutron scattering can be inferred from the tunneling data by
analyzing the derivatives of SIN and SIS conductances. We found that the derivative of the SIN
conductance diverges at $eV = \Delta + \Omega_{res}$ while the 
derivative of the SIS conductance
diverges at $eV = 2\Delta + \Omega_{res}$, where $\Delta$ is the maximum value
of the $d-$wave gap.

It is our pleasure to thank G. Blumberg, A. Finkel'stein and 
particularly J. Zasadzinski
 for useful conversations. 
The research was supported by NSF DMR-9979749.

\end{document}